# EVALUATION OF A LINGUAL INTERFACE AS A PASSIVE SURGICAL GUIDING SYSTEM

JOSE VAZQUEZ-BUENOSAIRES, YOHAN PAYAN AND JACQUES DEMONGEOT

Laboratoire TIMC, Equipe GMCAO, Fac. de Méd. de Grenoble, Domaine de la Merci, 38706,La Tronche Cedex, France

A general classification has been established for guiding systems in the field of Computer-Aided Surgery (CAS), based on the type of interaction (also called degree of passivity) between the human operator and the device [7]:

- *Passive systems*: the surgeon is totally responsible for the execution of the surgical action.
- *Active systems* realise part of the intervention autonomously and the operator supervises the task.
- *Semi-active systems* involve a combined action with the human operator for the complete realisation of the task.
- *Synergistic systems* allow the surgeon to have control of some degrees of freedom, while the device controls others.

These classes establish a list of qualitative factors according to the task- or to the user-oriented properties. Among task-oriented properties are:
The geometry complexity: as "free motion" or "navigation", "positioning", "trajectory following",
- "motion in a region" and "intra-cavity motions".
- The need for force control.
- The need for reaction to motions.
- The need to get very high position and force resolutions (for robotized systems) and sensitivity.

To another extend, user-oriented properties are:
- Safety
- The need to exploit the adaptability of the system [6].

In this paper, a surgical guiding system based on lingual stimulation is introduced, and evaluated in relation to the factors listed above. This system, called the Tongue Display Unit (TDU) was initially introduced and evaluated in the context of rehabilitation studies for blind people [1][2][5]. A small matrix of 12 by 12 electrodes is put in contact with the tongue surface, and a signal is sent to each electrode in order to stimulate the lingual mucous.

In this context of sensory substitution, the interface has demonstrated (1) its capacity to conveying meaningful information to the user and (2) the higher discrimination of the tongue in spatial acuity, compared to haptics interfaces [5]. The TDU is introduced here in a Computer-Aided Surgery framework, to per-operatively guide the surgeon gesture.

With the use of per-operative CAS systems, the surgeon is now even more confronted to problems of saturation by excessive multi-modal information. The aim of this work is to send back the information collected by the CAS system through one channel that is neither the visual nor the auditory channel, already highly saturated (patient, screens, videos, conversations, etc.).

As a first step, to validate the feasibility of the system, the information supplied to the TDU interface is the real-time measured position of a surgical instrument inside a volumetric model, in order to maintain the surgeon's gesture as close as possible to a pre-planned trajectory. This 3D rigid body is defined by 6 degrees of freedom (DOF) that correspond to the position and orientation of the ancillary. A coding has therefore to be chosen to transform the 6D-measured information into the 2D surfacic TDU interface.

## MATERIAL

### Introduction

The prototype system architecture uses a graphical user interface (GUI) to control the TDU interface and the optical localizer Polaris® (Northern Digital Inc.).

The idea is that the position of the ancillary is localized by the Polaris and compared with the planned trajectory. The measured difference between actual and planned trajectory is coded and sent to the TDU(Fig. 1):

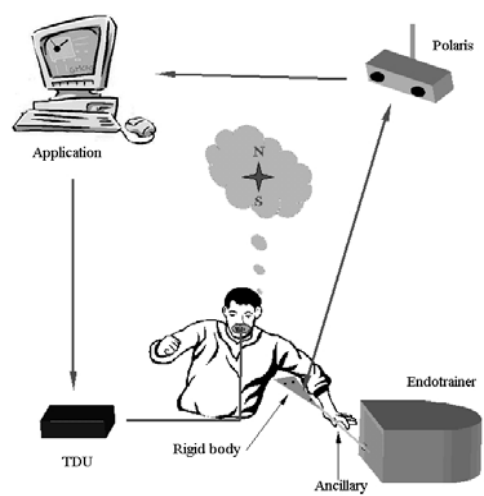

*Fig. 1:*
*Configuration of the system*

### 1. Graphical user interface

This application implements three important steps:
- *Calibration* of the Polaris and a surgical tool. Two points are calibrated (over the same axe) to track the tool in a 3D space ($Pt_1$, $Pt_2$), by using the pivot's method [4].
- *Preparation*: a virtual model, pre-acquired from scanner data, is built. The surgeon can therefore plan the "optimal" trajectory onto this virtual model.
- *Simulation*: This step integrates the planning and the calibration steps to evaluate the guiding process. For this, the tracking of the surgical tool is implemented and the measured position of the tool is coded and sent to the TDU interface. A phantom was used for this preliminary experiment.

## *2. TDU interface*

The TDU Interface consists in a thin strip of polyester material with a rectangular matrix of 144 gold-plated circular electrodes deposited by a photolithographic process. Each electrode separation's distance is 2.32 mm, and the 12x12-size array is 2.95 cm square [3]. The user puts the thin strip over his tongue to receive the electric stimuli. Before using this system, the user has to be trained, mainly to learn how the coding works.

## *3.Polaris® localizer*

The localizer tracks the position of two rigid bodies: one mounted on a surgical tool and the other one used as a reference, to calculate, in real time, the position and transformations of the two points previously calibrated. This information, corresponding to position and orientation of surgical instrument, is compared with the pre-planned trajectory to measure errors and is coded to use it as feedback information to the interfaceable strip.

## TRAJECTORY CODING

The code that was chosen as a first evaluation generates a simple representation based on two informations:
- the actual tool tip position, compared with the planned trajectory,
- the orientation of the tool,

and developed in two times:
- initially, the user is asked to fix the entry point in this case the movement

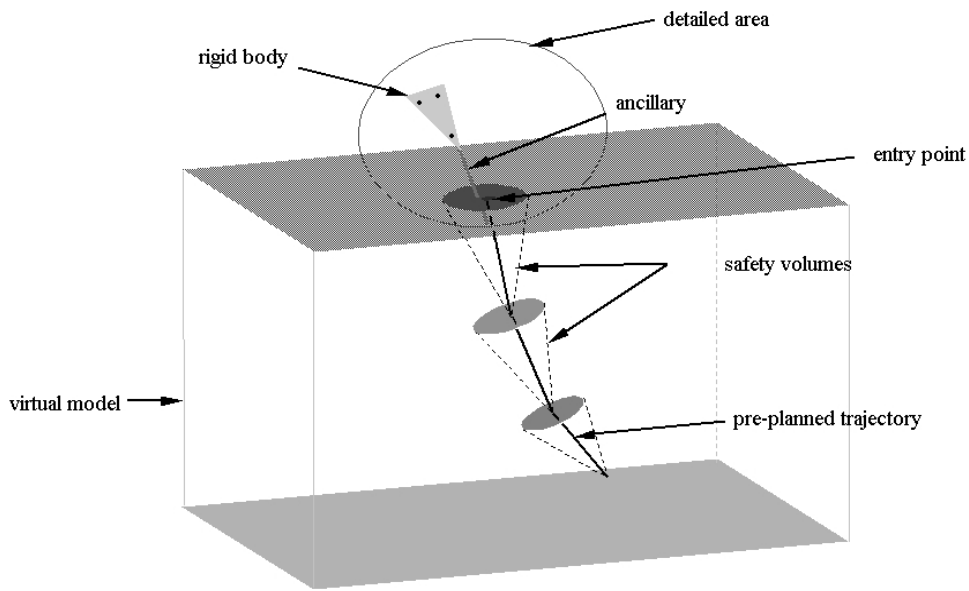

*Fig. 2:*
*Parameters of trajectory.*

represents displacement over a plane that is orthogonal to the 3D trajectory; the GUI therefore only supplies the stimuli that correspond to the new position of the tip,
- then the user has to follow the trajectory, with movements that have to be parallel to the trajectory direction; the GUI therefore only supplies the stimuli that correspond to the orientation of the tool (a direction parallel to the trajectory means no stimuli). At the same time, to maintain the tip displacement closer enough to the trajectory, a conical volume is specified, representing a virtual security confined zone. This virtual zone is coded onto the TDU interface by the mean of peripheral stimulation when the tool orientation differs from the planned trajectory direction.

In the example given in Fig. 3, the user has to establish the closer position to the trajectory entry point on the patient extend skin surface. For this, the actual position of the tip(Fig. 3a) is tracked by the Polaris, and sent to the TDU.

A bad tip position is represented by a square stimuli (formed by the activation of four electrodes) with a position from the centre proportional to error measured from tip to pre-planned entry point. A good tip position at the entry points means the activation of the TDU, in its central part (Fig. 3b).

Then, the orientation of the ancillary is measured and compared with the pre-planned trajectory orientation. The error is coded by the activation of peripheral pins closer to the higher part of ancillary (Fig. 3c). In this example, a given corner is active and the user has to avoid this direction to keep the right ancillary orientation (Fig. 3d).

The localizer tracks the position of two rigid bodies: one mounted on a surgical tool and the other one used as a reference, to calculate, in real time, the position and transformations of the two points previously calibrated. This information, corresponding to position and orientation of surgical instrument, is compared with the pre-planned trajectory to measure errors and is coded to use it as feedback information to the interfaceable strip.

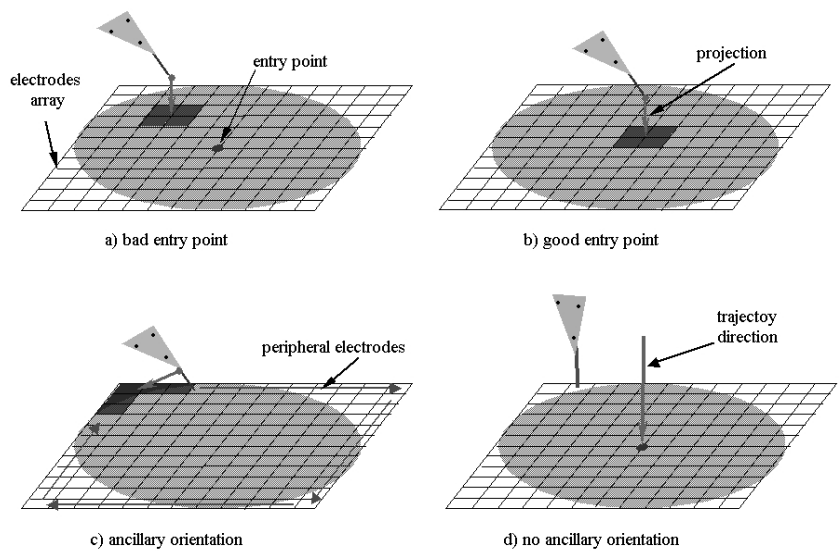

*Fig. 3 :*
*Detailed area describing :*
 *a), b): tip code and*
 *c), d): orientation code example.*

# FIRST QUALITATIVE TEST

To validate the effectiveness of this coding, five trajectories were assumed to reach the target. After a small training phase, five users were asked to follow as close as possible each of the simulated trajectories. Each user was asked to do six trials per trajectory. The first trajectory is perpendicular to the entry point surface. The second, third and fourth trajectories have respectively an approximately 70°, 30° and 10° deviation. Whereas all those trajectories were defined by only two points (the entry point and the target), the fifth trajectory was defined by three points, making therefore a curved trajectory.

In each test, the errors between actual trajectory and planned trajectory were measured. The mean time to recognise the entry point was of 15s. Table 1 presents the mean error, standard deviation, maximal error and the number of measured points acquired for each test. This number of points is correlated to the duration of the gesture.

# DISCUSSION

In this paper, a new system, the Tongue Display Unit was introduced to assist the guiding of a surgical gesture. This system seems to satisfy some of the quantitative factors in the general taxonomy of surgical guiding:

| Number of Trajectory | Total acquired points | Errors | test 1 | test 2 | test 3 | test 4 | test 5 | test 6 |
|---|---|---|---|---|---|---|---|---|
| 1 | 329 | Mean | 5.77 | 4.37 | 2.02 | 4.87 | 5.34 | 5.12 |
| 1 | 329 | Max | 10.66 | 8.96 | 5.20 | 13.46 | 12.53 | 17.63 |
| 1 | 329 | SD | 2.38 | 1.86 | 1.14 | 2.82 | 2.71 | 3.25 |
| 2 | 578 | Mean | 4.77 | 20.63 | 5.03 | 4.57 | 6.26 | 3.94 |
| 2 | 578 | Max | 11.03 | 66.66 | 10.47 | 9.35 | 12.95 | 5.78 |
| 2 | 578 | SD | 2.27 | 16.14 | 1.92 | 2.23 | 2.59 | 1.03 |
| 3 | 299 | Mean | 8.46 | 9.67 | 1.73 | 5.33 | 6.59 | 5.07 |
| 3 | 299 | Max | 15.82 | 21.57 | 4.49 | 8.64 | 12.37 | 9.71 |
| 3 | 299 | SD | 2.92 | 3.66 | 0.99 | 1.47 | 2.47 | 1.94 |
| 4 | 82 | Mean | 2.81 | 2.54 | 3.83 | 4.75 | 3.80 | 3.39 |
| 4 | 82 | Max | 4.20 | 5.13 | 6.66 | 11.35 | 8.53 | 7.86 |
| 4 | 82 | SD | 0.92 | 1.08 | 1.32 | 2.35 | 1.64 | 1.71 |
| 5 | 185 | Mean | 15.55 | 19.58 | 8.71 | 11.54 | 10.99 | 5.52 |
| 5 | 185 | Max | 37.98 | 44.60 | 17.44 | 26.75 | 23.45 | 13.83 |
| 5 | 185 | SD | 9.83 | 12.08 | 5.88 | 7.63 | 6.02 | 3.97 |

*Table 1:*
*Error measurements for each test and trajectory (in mm)*

-"free motion". An optical localizer is used to measure the position of a surgical tool,
-"Positioning" and "Trajectory following", is covered by the use of a lingual electro-stimulation, in order to inform the surgeon of the tool actual position and direction.
-"Motion in a region" is satisfied through the coding of a virtual cone around the target region.
The only qualitative factor that is not satisfied here is the force control, because classical haptic force feedbacks are replaced by a tactile feedback.

The first very qualitative results show that users, without any long-term training, are able to follow a simulated trajectory within a precision that is in mean below 5-mm. Those results seem to demonstrate the feasibility of the guiding system, but need of course to be more quantitatively validated. Subjects have to be trained and much more complex trajectories need to be evaluated.

## BIBLIOGRAPHY

**[1] BACH-Y-RITA P., COLLINS C.C., SAUNDERS F., WHITE B., SCADDEN L.**
*Vision substitution by tactile image projection*
Nature, 1969, 221: 963-964.

**[2] BACH-Y-RITA P.**
*Brain mechanisms in sensory substitution*
Academic Press, 1972: 182 pp.

**[3] BACH-Y-RITA P.**
*Sensory prostheses: tactile visual substitution systems; Conference: "The impeding paradigm shift in neurorehabilitation and remediation: The melding of basic research in neurosciences and behavioural science to produce advances in therapeutics"*
University of Alabama at Birmingham. 2001, July: 20-22.

**[4] LAVALEE S., CINQUIN P. AND TROCCAZ J.**
*Computer Integrated Surgery and therapy: State of the Art*
Contemporary Perspectives in Three-Dimensional Biomedical Imaging, 1997: 239-310.

**[5] SAMPAIO E., MARIS S., BACH-Y-RITA P.**
*Brain plasticity: 'visual' acuity of blind persons via the tongue*
Brain Research 2001, 908 : 204-207.

**[6] TAYLOR R. H. ET AL.**
*Augmentation of human precision in computer-integrated surgery*
Innovation and Technology in Biology and Medicine, 13 : 450-468

**[7] TROCCAZ J., PESHKIN M. AND DAVIES B.**
*Guiding systems for computer-assisted surgery: introducing synergistic devices and discussing the different approaches*
Medical Image Analysis. 1998, vol. 2: num. 2 : 101-119.